\newcommand{\ket}[1]{\ensuremath{\vert #1 \rangle}}

\documentclass[aps,prl,twocolumn,showpacs,preprintnumbers,amsmath,amssymb]{revtex4}
\usepackage{graphicx}
\usepackage{dcolumn}
\usepackage{mathrsfs}
\usepackage{bm}

\usepackage{amsfonts}
\usepackage{amssymb}
\usepackage{amsmath}
\usepackage{color}

\begin{document}

\preprint{}

\title{Electro-Optic Modulation of Single Photons}

\author{ Pavel Kolchin 
} \email{pkolchin@stanford.edu}
\author{Chinmay Belthangady}
\author{Shengwang Du}
\author{G.Y. Yin}
\author{S.E. Harris}

\address{
Edward L. Ginzton Laboratory, Stanford University, Stanford, California 94305, USA
}
\date{\today}

\begin{abstract} We use the Stokes photon of a biphoton pair to set the time origin for electro-optic modulation of the wave function of
the anti-Stokes photon thereby allowing arbitrary phase and
amplitude modulation.  We demonstrate conditional single-photon
wave functions composed of several pulses,  or instead, having
gaussian or exponential shapes.

\end{abstract}

\pacs{42.50.Gy, 32.80.Qk, 42.50.Ex, 42.65.Lm}
\maketitle

This letter demonstrates how single photons may be modulated so as
to produce photon wave functions whose amplitude and phase are
functions of time. The essential feature of this work is the use of one
photon of a biphoton pair that is generated by spontaneous
parametric down-conversion to establish the time origin for the
modulation of the second photon.  This is done by using
electromagnetically induced transparency and slow light to produce
 time-energy entangled biphotons with pulse lengths of several
hundred ns, and therefore, very long as compared to the temporal
resolution of single photon counting modules (about 40 ps). Once
the time origin is established, the photon waveform may be
modulated in the same manner as one modulates a classical pulse of
light. For example, the single-photon waveform may be phase,
frequency, amplitude, or even digitally modulated, with the
maximum modulation frequency limited by the resolution of the
detection of the first  photon.

\begin{figure}[htbp]
\begin{center}
\includegraphics [width=3.4 truein]{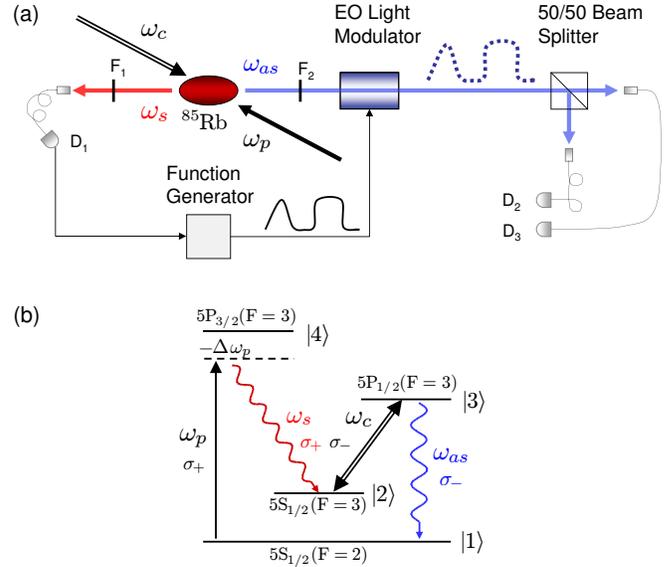}
\end{center}
\caption{(color online) (a) Schematic of paired photon generation
and conditional modulation. A Stokes photon detected by an SPCM
sets the time origin for shaping  the anti-Stokes photon with an
electro-optic modulator. To within the accuracy of the SPCM, this
allows shaping of both the amplitude and phase of the anti-Stokes
photon. (b) Energy level diagram of $\rm^{85}$Rb for paired photon
generation. } \label{fig:schematic}
\end{figure}

As shown in Fig.~\ref{fig:schematic}, we use cw pump and coupling
lasers to generate time-energy entangled pairs of Stokes and
anti-Stokes photons that propagate in opposite directions and are
collected in single mode fibers. The detection of a Stokes photon
at $D_{1}$ sets the time origin for firing the function generator
that drives the electro-optic modulator (EOM) that  modulates the
wave function of the anti-Stokes photon.  This latter photon is
incident on the beam splitter where it is detected by $D_{2}$ or
$D_{3}$.  As shown in the following, we generate single photons
whose modulated waveform is two rectangular pulses, is Gaussian or
is a time reversed exponential.

The method demonstrated in this letter might be used to optimally
load a single photon into an optical cavity~\cite{Zoller}, or
instead, to study the transient response of atoms to different
single photon waveforms. In the context of light-matter
interfaces, it may improve the efficiency of storage and retrieval
of single photons in atomic ensembles~\cite{Gorshkov}. For quantum
information applications, both amplitude and phase modulators
could be used to allow full control over the single photon
waveforms. For example, one could a construct a single photon
waveform that is a train of identical pulses with information
encoded into the relative phase difference between consecutive
pulses~\cite{Yamamoto}.

The generation of single photons with controlled waveforms has
been demonstrated earlier by using the techniques of cavity-QED,
i.e., by coupling a single trapped ion to a high Q cavity and
using an acousto-optic modulator to shape a pumping laser that is
tuned close to the resonant transition~\cite{Keller}.  In related
work Rempe and colleagues use a single Rb atom, again in a high
Q-cavity to generate single photons~\cite{Rempe}. The suggestions
and experiments of the Lukin and Kimble groups
~\cite{Lukin,Kimble} together with the cavity approach of
Vuletic~\cite{Vuletic} as well as the rapidly expanding work in
the category of quantum storage followed by single-photon
generation~\cite{Chou-Kimble,Eisaman-Lukin, Eisaman-LukinNature,
Kuzmich1,Kuzmich2, Laurat,Pan} offer other techniques for
generating conditional single-photons. More generally, the use of
one photon of a correlated pair as a trigger for a conditional
second photon is reviewed by Lounis and Orrit~\cite{Lounis}.

In this work we use the technique of Balic et al.~\cite{Balic,
Kolchin-theory, RaymondOoi} to generate biphotons in a backward
wave geometry where the length of the biphoton is determined by
the slow group velocity associated with EIT. We first summarize
the properties of this type of biphoton light source as recently
developed by Du and colleagues~\cite{Du}. This source uses  a
two-dimensional MOT with a cylindrical atomic cloud with a length
of 1.7 cm, an aspect ratio of 25, and an estimated dephasing rate
of the non-allowed $\ket{1}\rightarrow\ket{2}$ transition of
$\gamma_{12} =0.01\  \Gamma_{3}$, where $\Gamma_{3} = 6$~MHz is
the spontaneous decay rate out of the state $\ket{3}$. This source
may be run at an optical depth in the range of 10 to 60 resulting
in group velocities at the anti-Stokes wavelength of $3 \times
10^5$ to $2 \times 10^4$ m/s. This allows the generation of
biphotons with temporal lengths that can be varied over the range
of 50 to 900 $\rm{ns}$ and have estimated linewidths as small as
0.75 MHz. The generation rate of paired photons is about $2
\times10^5$ pairs per second, with a ratio of paired to
anti-Stokes photons of about $75\%$. These results are in
agreement with theory~\cite{Balic, Kolchin-theory} and no longer
exhibit the discrepancy reported in~\cite{Balic,Kolchin,Du}.
\begin{figure}[htbp]
\begin{center}
\includegraphics*[width=3.4 truein]{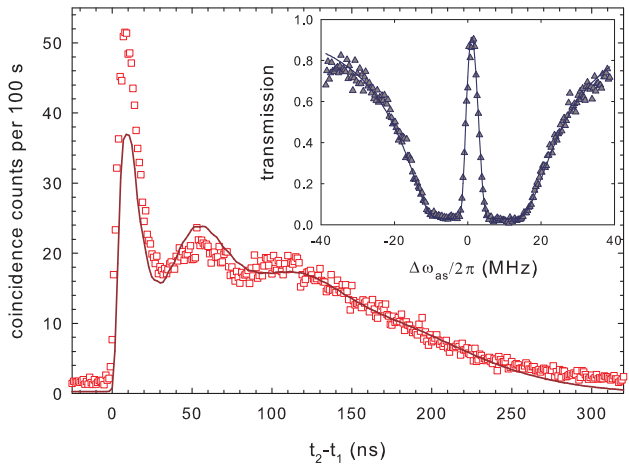}
\end{center}
\caption{(color online) No modulation; $D_1$-$D_2$ coincidence
counts in a 1 ns bin as a function of the relative delay at an
optical depth of 40. The inset shows the EIT transmission profile
for the anti-Stokes photon (see text for  further detail).}
\label{fig:nomod}
\end{figure}

Figure~\ref{fig:nomod} shows the coincidence count rate of
detectors $D_{1}$ and $D_{2}$  without modulation at an optical
depth of 40.  In Fig.~\ref{fig:nomod}, the Rabi frequencies of the
pump and coupling lasers are $\Omega_{c} = 2.2~  \Gamma_{3}$ and
$\Omega_{p} = 0.59~ \Gamma_{3}$, and the detuning of the pump from
resonance, $\Delta\omega_p =24.4~  \Gamma_{3}$. With $\tau$ as the
relative delay between Stokes and anti-Stokes detection times
$t_1$ and $t_2$ respectively, and $T_b$ as the bin width, the
Glauber correlation function is related to the coincidence rate,
$R_c (\tau)$  by $G^{(2)}(\tau)= R_c(\tau)/T_b$. This correlation
function may, in turn, be expressed in terms of the square of the
absolute value of the biphoton wave function $\Psi(t_1, t_1+\tau)$
and the rate of generation of Stokes ($R_s$) and anti-Stokes
($R_{as}$) photons as $G^{(2)}(\tau)=|\Psi(t_1, t_1+\tau)|^2+R_s
R_{as}$.

Two features of the correlation function in Fig.~\ref{fig:nomod}
are: a  mean width that is equal to the group delay time, as
caused by EIT, of the anti-Stokes photon relative to the Stokes
photon, and a sharp leading edge spike. This spike is a
Sommerfeld-Brillouin precursor and has an approximate width that
is equal to the opacity width of the EIT profile in the optically
thick medium~\cite{Brillouin,Du-precursor}. Figure~\ref{fig:nomod}
also shows the theoretically computed correlation function. In
order to account for experimental factors the theoretical curve is
multiplied by a factor of $\eta = 8.8\times 10^{-4}$. This scaling
factor takes into account the $10\%$ duty cycle, $57\%$
transmission at the beamsplitter port, additional losses of about
$20\%$ in the beamsplitter, Stokes and anti-Stokes filters
efficiencies of $48\%$ and $42\%$ respectively, fiber to fiber
coupling of $75\%$, a detector efficiency of $50\%$ and EO~light
modulator insertion loss of $50\%$. Agreement between theory and
experiment is very good. This agreement is important because it
distinguishes the biphoton wavefunction from background counts
which may also be modulated.

The electro-optic amplitude modulator consists of  phase
modulators in both arms of a Mach-Zehnder (MZ)
interferometer~\cite{Keang}. The degree of phase control in both
arms depends on the type of the modulator. We use a z-cut
modulator that requires $V_{\pi} = 1.3$~volts to cause the $\pi$
phase shift required to go from minimum to maximum transmission
and can be operated at a maximum frequency of 10~GHz. One port of
the output beam splitter of the MZ interferometer is terminated so
that the portion of the photon wave function that is not
transmitted is lost. In general, if a Stokes photon is detected at
time $t_1$, and the modulator is activated conditioned on this
detection then, in the Heisenberg picture, the anti-Stokes
operator at the output of the modulator is related to the input
operator by $\hat{a}_{out}(t_2) =
\int{g(t_2,t^\prime_2)\hat{a}_{in}(t^\prime_2)dt^\prime_2}$. If
there are no dispersive elements, then to within an unimportant
phase factor we may write
$\hat{a}_{out}(t_2)=m(\tau)\hat{a}_{in}(t_2)$ . The correlation
function in the presence of the modulator is related to that in
the absence of the modulator by
\begin{equation}
G_m^{(2)}(\tau)=|m(\tau)|^{2} G^{(2)}(\tau)
\end{equation}
With the biphoton wavefunction given by $\Psi (t_1, t_1+\tau)$ the
modulated (conditional) single photon wavefunction is $m(\tau)\Psi
(t_1, t_1+\tau)$. We adjust the bias voltage at the input of
modulator so that the output of the modulator $m(\tau)$ is related
to the input voltage $V(\tau)$ by $m(\tau) = \sin
\left[\phi(\tau)\right] \exp \left[i \alpha \phi(\tau) \right]$,
where $\phi(\tau) = \pi V(\tau)/(2 V_{\pi})$, and $\alpha$ is a
phase modulation parameter. For a z-cut amplitude modulator as
used here $\alpha = 0.75$, but may be eliminated by using an x-cut
modulator.

In somewhat more detail, the detection of a Stokes photon by
detector $D_1$ (Perkin-Elmer SPCM-AQR-13) triggers both the start
input of a time-to-digital converter (FAST Comtec TDC 7886S) as
well as a function generator (Agilent 33250A-80MHz) which
generates the desired modulation signal. The function generator
along with auxiliary logic gates has an electronic delay of about
190~ns. To partially compensate for this delay the anti-Stokes
photon is optically delayed by 140~ns using a 30~m long single
mode polarization maintaining fiber.  The anti-Stokes beam is then
passed through the EO modulator (EOSPACE Inc.) driven by the
output of the function generator. Verification of the single
photon nature of the modulated anti-Stokes photon is done using a
50-50 beam splitter and detectors $D_2$ and $D_3$, which are
connected to the stop inputs of the time-to-digital converter.
Coincidence counts are binned into histograms with 1~ns bin width
and plotted as a function of the difference between the arrival
times of the Stokes and anti-Stokes photons.
\begin{figure}[htbp]
\begin{center}
\includegraphics*[width=3.4 truein]{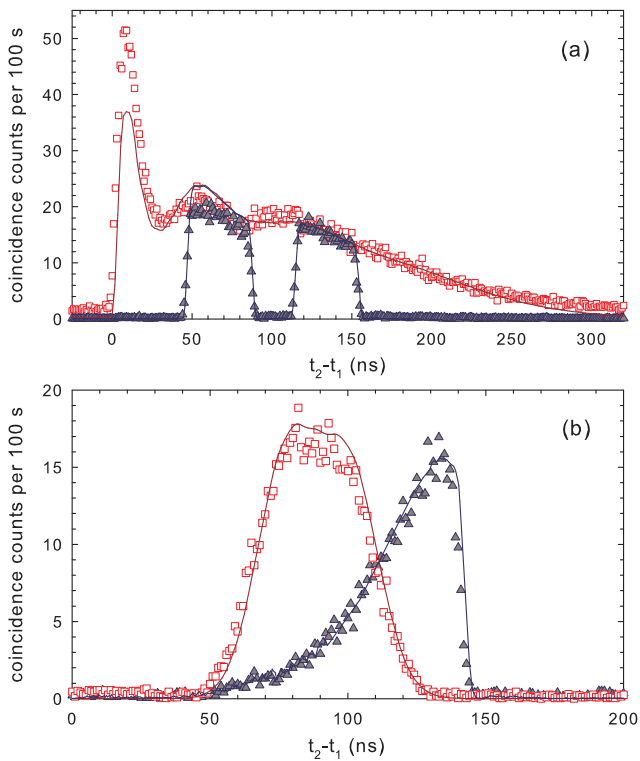}
\end{center}
\caption{(color online) $D_1$-$D_2$ coincidence counts in a 1~ns
bin as a function of the delay between Stokes and anti-Stokes photons. (a)
Modulated ($\bigtriangleup$) and unmodulated ($\square$)
waveforms. (b)  Waveforms with Gaussian ($\square$) and rising
exponential ($\bigtriangleup$) shapes. The experimental data
($\square$, $\bigtriangleup$) were collected over 2000~s. The solid
curves are plotted from theory. } \label{fig:shapes}
\end{figure}

As a first example, we use two rectangular pulses as the
modulation signal. Figure~\ref{fig:shapes}(a) shows coincidence
counts between detectors $D_1$ and $D_2$ with and without
modulation. In the latter case the EO~light modulator is set to
maximum transmission. Of importance, there is no vertical scaling
between the modulated and non-modulated waveforms. Coincidence
histograms between detectors $D_1$ and $D_3$ (not shown) have
similar shapes.  In Fig.~\ref{fig:shapes}(b), we show modulated
photons with two different waveforms. In the first case we drive
the modulator with a gaussian pulse.  In the second case we design
the function generator output waveform so as to compensate for the
functional(or nonlinear) distortion in such a way that the output
of the modulator is an exact rising exponential. The solid lines
show theoretical curves which are calculated by using the scope
traces of the output voltage of the function generator as
$V(\tau)$.

We define the retrieval efficiency as the probability to generate
a single anti-Stokes photon on the condition that the paired
Stokes photon is detected, $\mathcal{E}_R = R_p/R_s$. Here, $R_p$
is the total paired rate, calculated from the area under the
correlation curves, measured at detectors $D_1, D_2$ and
$D_1,D_3$, minus the uncorrelated background floor. When the
modulator is fully open [Fig.~\ref{fig:nomod}] we measure
$\mathcal{E}_R= 3.5\%$. When losses at the beamsplitter,
modulator, filters, fiber to fiber coupling and detector
efficiency in the anti-Stokes path are backed out, this corresponds
to a retrieval efficiency of $55\%$. With the modulator present the
retrieval efficiency is waveform dependent.  For the examples of
two square pulses, rising exponential and  gaussian waveforms,
the retrieval efficiency is  $\mathcal{E}_R =1.3\%$, $0.61\%$ and
$0.9\%$, respectively. If no losses were present, these
efficiencies would be $21\%$, $9.4\%$ and $11.2\%$.
\begin{figure}[htbp]
\begin{center}
\includegraphics*[width=3.2 truein]{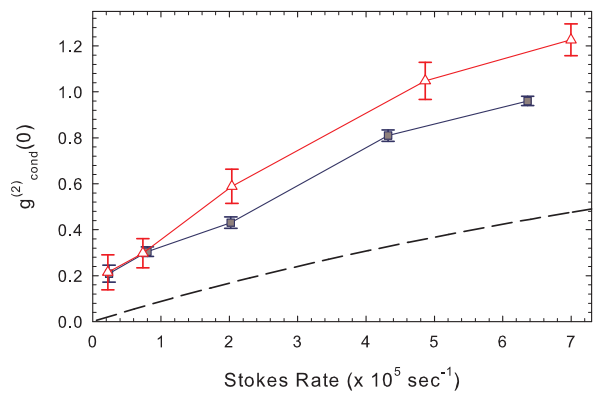}
\end{center}
\caption{(color online) Conditional three-fold correlation
function $g^{(2)}_{cond} (0)$ as a function of the Stokes rate for
unmodulated ($\Box$) and modulated ($\bigtriangleup$)
single-photon generation. The dashed curve shows the theoretical
limit for $g^{(2)}_{cond} (0)$ in the absence of excess light scattering (see text).} \label{fig:condG2fig}
\end{figure}

Since single photons incident on a beamsplitter must go into one
output port or the other, in the ideal case where there are no two-photon
events and there is no excess scattered light, we would expect no three-fold
coincidences at the detectors. A measure of the quality of
heralded single photons that quantifies suppression of two photon
events is given by the conditional correlation function~\cite{Grangier}:
\begin{equation}
\begin{split}
g^{(2)}_{cond} (0) &= \frac{N_{123} N_{1}}{N_{12}
N_{13}}.\label{eq:G2cond}
\end{split}
\end{equation}
Here $N_1$ is the number of the Stokes counts at $D_1$, $N_{12},
N_{13}$ are the number of two-fold coincidence counts at detectors
$D_1$, $D_2$ and $D_1$, $D_3$ respectively within a time window
$T_c$, and $N_{123}$ is the number of three-fold coincidence counts
within the same time window.

In Figure~\ref{fig:condG2fig}  triangles and squares show measured
$g^{(2)}_{cond} (0)$ versus Stokes rate with and without
modulation. The modulation is done with the same signal as in
Fig.~\ref{fig:shapes}(a). We set $T_c$ equal to the nominal length of
the unmodulated biphoton ($285$~ns). At a Stokes rate of $2.2
\times 10^4$~sec$^{-1}$ which corresponds to $\Omega_p =
0.26~\Gamma_3$, we obtain $g^{(2)}_{cond} (0) = 0.2 \pm 0.04$ and
$g^{(2)}_{cond} (0) = 0.21 \pm 0.07$ for the unmodulated and
modulated waveforms respectively. The fact that the measured
$g^{(2)}_{cond} (0)$ is less than $0.5$, (the limiting value
for a two photon Fock state), is indicative of the
near-single photon character of the light source.

Because there is a small probability for the parametric down conversion process to generate multiple pairs of biphotons, even in the absence of spurious light scattering, the conditional correlation function is not zero. The dashed curve In Fig.~\ref{fig:condG2fig} shows the theoretical
prediction for the conditional correlation function that results from such multiple scattering events. Because of light scattering from both the pump and coupling lasers, the experimental curves lie above this limiting value.

We describe two control experiments: In the first
we remove the 30~m long optical fiber so as to modulate the
uncorrelated background noise in the tail of the correlation
function. Here, we measure $g^{(2)}_{cond} (0) = 1.2$. In
the second experiment we apply modulation at random times, using
an external 10 MHz digital signal as a trigger for the function
generator. As expected, we observe a reduced rate of paired counts and no change in the shape of the
correlation function.

This Letter has demonstrated a technique for using one photon of a biphoton pair to set the time origin and to trigger an electro-optic modulator to shape the waveform of the second photon.  The importance of the electro-optic method is its speed and ability to modulate phase as well as amplitude. The technique provides the technology for studying the response of atoms to shaped single photon waveforms on a time-scale comparable to the natural linewidth.

The authors acknowledge helpful discussions with Tony Siegman,
Steve Sensarn and Irfan Ali-Khan. The work was supported by the
U.S. Air Force Office of Scientific Research, the U.S. Army
Research Office and the Defense Advanced Research Projects Agency.

\end{document}